\documentclass[sigplan,screen]{acmart}
\usepackage{booktabs}
\usepackage[utf8]{inputenc}
\usepackage[T1]{fontenc}
\usepackage{todonotes}
\usepackage{pgfplots}
\usepackage{array}
\usepackage{multirow}
\usepackage{graphicx}

\setcopyright{rightsretained}

\begin{document}

\acmPrice{}
\acmDOI{10.1145/3372885.3373827}
\acmYear{2020}
\copyrightyear{2020}
\acmISBN{978-1-4503-7097-4/20/01}
\acmConference[CPP '20]{Proceedings of the 9th ACM SIGPLAN International Conference on Certified Programs and Proofs}{January 20--21, 2020}{New Orleans, LA, USA}
\acmBooktitle{Proceedings of the 9th ACM SIGPLAN International Conference on Certified Programs and Proofs (CPP '20), January 20--21, 2020, New Orleans, LA, USA}

\title[Exploration of Neural Machine Translation in Autoformalization of Mathematics in Mizar]{Exploration of Neural Machine Translation in Autoformalization of Mathematics in Mizar}

\author{Qingxiang Wang}
\affiliation{
  \department{Department of Computer Science}             
  \institution{University of Innsbruck}           
  \country{Austria}                   
}
\email{shawn.wangqingxiang@gmail.com}
\author{Chad Brown}
\affiliation{
  \department{CIIRC}             
  \institution{Czech Technical University in Prague}           
  \country{Czech Republic}                   
}
\author{Cezary Kaliszyk}
\affiliation{
  \department{Department of Computer Science}             
  \institution{University of Innsbruck and University of Warsaw}           
  \country{Austria}                   
}
\email{cezary.kaliszyk@uibk.ac.at}         
\author{Josef Urban}
\affiliation{
  \department{CIIRC}             
  \institution{Czech Technical University in Prague}           
  \country{Czech Republic}                   
}
\email{josef.urban@gmail.com}

\begin{abstract}
In this paper we share several experiments
  trying to automatically translate informal mathematics into formal mathematics.
In our context informal mathematics refers to human-written mathematical sentences in the LaTeX format;
and formal mathematics refers to statements in the Mizar language.
We conducted our experiments against three established neural network-based machine translation models that are known to deliver competitive results on translating between natural languages.
To train these models we also prepared four informal-to-formal datasets.
We compare and analyze our results according to whether the model is supervised or unsupervised.
In order to augment the data available for auto-formalization and improve the
results, we develop a custom type-elaboration mechanism and integrate it in the
supervised translation.
\end{abstract}



\maketitle

\section{Introduction}
Formalization of all existing mathematics from textbooks and literature has been a long-term dream for researchers in interactive theorem proving (ITP)~\cite{qed-manifesto}.
For decades we have witnessed the growth of many ITP proof assistant libraries, some of which can now cover quite a proportion of mathematics.
However, despite such growth, at the current stage formalizations are still done manually and by a small community of domain experts.
This makes the speed of formalization much slower than the speed of increase of mathematics.
In order to work toward the vision of formalizing all existing mathematics, we need to figure out ways to automate the formalization process or parts of it.
We coin this research effort as \textit{autoformalization}.

Autoformalization is generally considered as a very challenging (or even impossible) task.
In addition to all the technical issues facing the ITP community, an autoformalization pipeline needs to capture the human thinking process that enables translation from informal mathematics texts into formal pieces of code in a proof assistant.
Therefore, some people argue that implementing such an autoformalization pipeline would amount to implementing a program that is capable of achieving sophisticated natural language understanding and mathematical reasoning.
We believe that the above holistic view can be simplified if we first approach the problem by just considering it as a language translation problem, i.e., given an informal mathematical sentence, we translate it into a formal sentence that has the potential to be further formalized into a proof assistant.
If such a translation tool were available, we would be able to preprocess the vast mathematical materials and come up with their formal versions that can be used for later steps of formalization.

By simplifying and considering formalization as a language translation problem, we can employ state-of-the-art machine translation tools and adapt them to our purpose.
Currently, the best machine translation models are all based on neural networks~\cite{FosterVUMKFJSWB18}.
As we are still in the exploration phase, in this paper we will not cover detailed engineering aspects of implementing neural network models but be satisfied with a high-level understanding of them, which we will introduce in Section \ref{s:nn}.
To simplify our engineering effort we will use three established neural machine translation models: 

1) The neural translation model (\textbf{NMT}) by Luong et al.~\cite{luong17} is a well documented model developed at Google based on the Tensorflow~\cite{tensorflow2015-whitepaper} framework.
It is an encoder-decoder architecture with a flexible range of configurable hyper-parameters.
The default configuration is a two-layer recurrent neural network with LSTM cells~\cite{lstm}.
Wang et al.~\cite{qwckju-cicm18} have conducted extensive experiments with this model in particular with hyper-parameter configurations suitable for autoformalization.
In this paper we extend the NMT with a type elaboration mechanism.

2) The unsupervised translation models (\textbf{UNMT}) based on Lample et al.~\cite{lample2018phrase} and Artetxe et al.~\cite{DBLP:journals/corr/abs-1710-11041} attempt to find a mapping between two languages
without any initial alignment data. Both models are implemented in the PyTorch~\cite{pytorch} framework. As both models share the same idea of transforming an unsupervised learning problem into a series of supervised learning problems, we only perform experiments with the first one, training it with with non-aligned autoformalization data.
UNMT can be configured to use both the NMT architecture as in 1) or the transformer architecture~\cite{vaswani2017}.
We will illustrate both ideas in Section \ref{s:nn}.

3) Lample et al.'s~\cite{lample2019cross} cross-lingual pretraining model (\textbf{XLM}) is a multi-task model. We only focus on its translation capability and treat it as an improved version of the UNMT model.
In UNMT a technique called word2vec~\cite{word2vec} is used to obtain vector representation of word tokens.
This vector representation is kept fixed during the whole training and evaluation process.
In XLM, a new technique
by Devlin at al.~\cite{bert} called \textit{pretraining} is adapted to directly obtain initial vector representations at both sentence level and word token level.
From these initial representations unsupervised learning can be performed to fine-tune the model.
Pretraining has been shown by the natural language processing (NLP) community to provide better results in many experimental benchmarks.
In this paper we will also see its improvement on our informal-to-formal datasets comparing to the original UNMT model.

The above models require training and testing data in the form of collection of sentences, i.e., a text corpus.
For the translation task we need a source corpus and a target corpus (in our case an \textit{informal mathematics} corpus and a \textit{formal mathematics} corpus).
Depending on whether the sentences in both corpora are aligned or not, we can employ the supervised model as in 1) or the unsupervised models as in 2) or 3).
To further clarify our research objective, in this paper we regard informal mathematics corpus as sentences in the LaTeX format, and formal mathematics corpus as statements in the Mizar language.
We prepared four datasets to test the above models.
The first three are LaTeX --- Mizar corpora with the informal part in increasing corpus size and resemblance to real natural language, while the last one corresponds to an ATP translation:

\begin{enumerate}
  \item Synthetic LaTeX --- Mizar dataset;
  \item ProofWiki --- Mizar topology dataset;
  \item ProofWiki --- Mizar full dataset;
  \item Mizar --- TPTP dataset.
\end{enumerate}

In Section \ref{s:data} we explain our decision to use Mizar as the formal corpus and share with the reader its ensuing strengths and weaknesses.
We also give a detailed account of the four datasets, explaining how they are prepared and giving a few examples.

In Section \ref{s:nn} we first briefly explain the background of neural machine translation and cover works that are related to the three models we used.
Then we give a high-level introduction to the three models to the point where we can apply them to our datasets.

In Section \ref{s:exp} we show our preliminary experimental results for training the translation models with our datasets.
We evaluate the translation quality by several metrics: the BLEU rate and perplexity provided by the models, the distribution of edit distances, as well as visual comparison.
Based on these we try to analyze their performance and point out the limitations of our current experiments.

In Section \ref{s:elab} a separate experiment is performed that incorporates a type elaboration mechanism with the supervised NMT model.
We introduce our motivation for this experiment and then describe the type elaboration algorithm on pattern formulas.

There are few related works to autoformalization of mathematics.
We discuss them in Section \ref{s:related} and conclude our paper in Section \ref{s:concl}.

\section{Informal-To-Formal Datasets}\label{s:data}

The choice of LaTeX sentences as informal corpus is justified by the fact that most of the mathematics literature nowadays are written in this format.
The choice of Mizar statements as formal corpus needs more explanation.

First of all, as our objective is to leverage data-driven machine learning models,
our chosen proof assistant
must already have
a large database of mathematical knowledge from which we can prepare a formal corpus.
Typical natural language corpus consists of millions of lines of sentences.
By comparison, none of the existing proof assistants can provide a corpus of this size.
Among all proof assistants, Mizar has more than 55 thousand theorem statements and more than 3 million lines of code in its Mizar Mathematical Library (MML)~\cite{BancerekBGKMNP18}.
This is the largest formal library that the ITP community can provide that has a comparable size to a natural language corpus. The only contender is the Isabelle library and its  Archive of Formal Proofs~\cite{BlanchetteHMN15}, which has more formalized computer science knowledge,
but less mathematics that can be overlayed with LaTeX texts.
It is mainly because of the library size that we chose Mizar as our formal language.

Secondly, although it is still early to discuss foundational issues, we believe that the formal mathematics we are considering  needs to be built on a formal theory that is capable to conveniently describe all contemporary mathematics.
Indeed there are areas of mathematics (e.g., certain branches of set theory) where one can always formulate a statement that cannot be described in a fixed semantics, but for the purpose of formalizing mainstream mathematics the underlying formal theory needs to be at least able to describe large categorical notions since they are now pervasive in many areas of mathematics.
This rules out several HOL-based systems, as from a denotational semantics point-of-view all non-polymorphic types are interpreted as members of a single fixed universe whose existence can be proved from ZFC, so the theory of HOL is weaker than the theory of ZFC~\cite{intro2HOL}, and ZFC is not enough for describing large categories without introducing the notion of class.
On the other hand, Mizar is based on Tarski-Grothendieck set theory.
The existence of universes is guaranteed by the Tarski A axiom~\cite{Trybulec89tarskigrothendieck}, therefore there are no major issues in describing large categories.

There are also minor reasons that contributed to our decision to use Mizar:
We want our formal language to be human readable so it can be easier for us to evaluate translation quality;
We want to minimize unicode processing and the MML is ASCII-based;
We also want the proof assistant library not to have a drastic syntactic difference between the theory language and its meta-language so that we can extract the maximum amount of information to formulate our corpus.

The above reasons for choosing Mizar do not mean that we think Mizar is better than other proof assistants.
What we want to claim is that we think Mizar is currently the most \textit{expedient} for us to test the effectiveness of neural machine translation models.
We also do not claim that Mizar will be the formal language in a potential auto-formalization pipeline.
Except the size of MML, all other reasons for choosing Mizar are simple and speculative.
In fact, by choosing a proof assistant not based on type theory, we lose the opportunity to easily use automation later on to possibly further improve the quality of translation, which is why in Section \ref{s:elab} we need to use a different Mizar --- TPTP dataset in order to test our type elaboration mechanism for data augmentation in the supervised learning setting.

\subsection{The Synthetic LaTeX --- Mizar Dataset}

By using existing tools it is possible to extract logical information from the MML.
Urban~\cite{mptpjar,mptp0.2} proposed a toolchain that can translate the MML into first-order proof obligations in the TPTP format~\cite{tptp94}.
These proof obligations can then be used by resolution-based first-order automated theorem provers (ATP).
The Mizar language is not purely first-order, its ability to form schemes of theorems as well as its soft-typing mechanism in formulation of statements make it possible to extract higher-order information from the MML.
This was verified by Brown et al.~\cite{brownhomml} and the translated higher-order TPTP proof obligations were subsequently tested by a resolution-based higher-order ATP~\cite{brownsatallax}.

However, to prepare a LaTeX-to-Mizar dataset, techniques and experience gained from previous efforts cannot be directly applied.
Although there are abundant mathematics literature in the LaTeX format that are freely available on the internet, parsing natural language to obtain logical information is known to be difficult.
In the NLP community, preparing aligned data usually requires intensive manual labor, iterative data cleansing and huge computing resources.
As a result, at the initial phase of our exploration it would be too costly for us if we went along this direction of directly facing the massive data.

To temporarily circumvent this problem, a possible approach is to generate informal sentences directly from the formal sentences we already have.
Such an \textit{informalization} approach has been adopted by Kaliszyk et al.~\cite{KaliszykUV15,KaliszykUV17I} to obtain aligned informal-to-formal corpora from the formal proof of the Kepler
conjecture~\cite{flyspeck} in HOL Light and the MML.
In those works informal corpora were made by merging overloaded symbols, deleting symbol prefixes, removing necessary parentheses, etc.
However, this approach can only generate \textit{ambiguated} informal sentences from formal sentences and they are not in the LaTeX format.

\begin{table}[h!]
\renewcommand{\arraystretch}{1.2}
\caption{The synthetic LaTeX --- Mizar dataset}\label{t:synthetic}
\begin{tabular}{m{0.65\columnwidth}<{\raggedright\arraybackslash}m{0.25\columnwidth}<{\raggedleft\arraybackslash}}
\toprule
Total sentence pairs & 1056478 \\
Unique sentence pairs & 605232 \\
Shortest Mizar sentence & 3 tokens \\
Shortest LaTeX sentence & 3 tokens \\
Longest Mizar sentence & 1526 tokens \\
Longest LaTeX sentence & 19012 tokens \\
\bottomrule
\end{tabular}
\end{table}

Fortunately, over the last decades the Mizar group has developed a tool that can be adapted to our purpose:
To facilitate the publishing of Mizar's journal \textit{Formalized Mathematics}, Bancerek et al.~\cite{bancerekmizar,bancerek2006automatic} developed a tool that can translate theorems and top-level proof statements from a Mizar article into artificial LaTeX sentences.
This tool first specifies translation patterns for all the Mizar constructors (e.g. functors, predicates, attributes, modes and structures) that are involved in that Mizar article.
Then it uses a combination of algorithms to handle the use of parenthesis, the regrouping of conjunctive formulas, grammatical correctness as well as other typesetting issues.
Finally, necessary style macros are added to polish the output rendering.

Comparing to other ITP-journals such as Isabelle's \textit{Archive of Formal Proofs}, Mizar's \textit{Formalized Mathematics} blurs the distinction between side remarks and formal code.
The artificial LaTeX sentences are quite readable and the whole translated article has an outlook that is similar to a human-written article.
We find these features appealing and spot the significance of Bancerek's tool for generating informal-to-formal corpora.

\begin{table*}[!htb]
\renewcommand{\arraystretch}{1.2}
\caption{Mizar vs. Synthetic LaTeX}\label{t:syntheticex}
\begin{tabular}{m{0.06\linewidth}<{\raggedright\arraybackslash}m{0.9\linewidth}<{\raggedright\arraybackslash}}
\toprule
Mizar & \texttt{\small cluster reflexive -> complete for 1 -element RelStr ;} \\
LaTeX & One can verify that every 1-element relational structure which is reflexive is also complete . \\
\LaTeX & One can verify that every 1-element relational structure which is reflexive is also complete . \\
\cmidrule{2-2}
Mizar & \texttt{\small let T be RelStr ;} \\
LaTeX & Let \$ T \$ be a relational structure . \\
\LaTeX & Let $ T $ be a relational structure . \\
\cmidrule{2-2}
Mizar & \texttt{\small mode type of T is Element of T ;} \\
LaTeX & \{ A type of \$ T \$ \} is an element of \$ T \$ . \\
\LaTeX & { A type of $ T $ } is an element of $ T $ . \\
\cmidrule{2-2}
Mizar & \texttt{\small attr T is Noetherian means : Def1 : the InternalRel of T is co-well\_founded ;} \\
LaTeX & We say that \{ \$ T \$ is Noetherian \} if and only if ( Def . 1 ) the internal relation of \$ T \$ is reversely well founded . \\
\LaTeX & We say that { $ T $ is Noetherian } if and only if ( Def . 1 ) the internal relation of $ T $ is reversely well founded . \\
\cmidrule{2-2}
Mizar & \texttt{\small a ast ( b ast t ) <= b ast t ;} \\
LaTeX & \$ a \textbackslash ast ( b \textbackslash ast t ) \textbackslash leq b \textbackslash ast t \$ . \\
\LaTeX & $ a \ast ( b \ast t ) \leq b \ast t $ . \\
\cmidrule{2-2}
Mizar & \texttt{\small 'not' Ex ( Ex ( a , A , G ) , B , G ) = All ( 'not' Ex ( a , B , G ) , A , G ) ;} \\
LaTeX & \$ \textbackslash neg \{ \textbackslash exists \_ \{ \textbackslash exists \_ \{ a , A \} G , B \} \} G = \{ \textbackslash  forall \_ \{ \textbackslash neg \{ \textbackslash exists \_ \{ a , B \} \} G , A \} \} G \$ . \\
\LaTeX & $ \neg { \exists _ { \exists _ { a , A } G , B } } G = { \forall _ { \neg { \exists _ { a , B } } G , A } } G $ . \\
\bottomrule
\end{tabular}
\end{table*}

Bancerek's original tool could only translate Mizar abstracts, i.e., top-level theorem statements, definitions and registrations.
In 2018, based on the early work of Urban~\cite{DBLP:conf/mkm/Urban05}, the Mizar group extended Bancerek's translation tool and applied it to all the Mizar articles~\cite{DBLP:conf/mkm/BancerekNU18}.
To generate aligned LaTeX --- Mizar sentences, this new tool keeps track of the offsets of Mizar statements throughout the intermediate steps, so the final LaTeX output can be traced back to its corresponding Mizar statement.
We use this new tool to generate an aligned LaTeX --- Mizar corpus of one million sentence pairs.
Since the LaTeX corpus is artificially generated from Mizar corpus instead of collected from human-written LaTeX sentences, we call this dataset the \textit{synthetic} dataset. Some statistics of this dataset are presented in Table~\ref{t:synthetic} and example sentences are presented in Table~\ref{t:syntheticex}.

As stated previously, obtaining aligned informal corpus from formal corpus algorithmically is only a temporary workaround to the data collection problem.
Since all the translation techniques involved are deterministic, although the generated LaTeX sentences are grammatically correct, they usually look artificial and vocabulary used is smaller than in human-written sentences.
This weakens the generalization capability of a translation model trained by only the synthetic dataset.

\subsection{The ProofWiki --- Mizar Datasets}

Translation models based on supervised learning require high-quality aligned corpora between source and target languages.
Unsupervised learning claims to get rid of aligned corpora, so that only monolingual corpora are needed for training.
However, at least for the two unsupervised models we are considering, a smaller aligned corpus for model validation and testing is still required.
In addition, to maximize the translation quality, both languages from the aligned corpus need to come from the same topic realm where the monolingual corpora are prepared.

In UNMT, the authors used English and French news articles in WMT'14 dataset as monolingual corpora, and parallel English-to-French news translations in WMT'17 dataset for model evaluation.
Since both are news articles, the linguistic variety and the vocabulary used are at the same range.
To adapt this model for our purpose, we also need to have sizable LaTeX and Mizar corpora with a subset of the two corpora aligned.
A simple solution to this requirement is to just use the previous synthetic dataset, keeping the alignment of a subset of it and forgetting the alignment of the rest.
However, in order for us to cope with the limitations of using the synthetic dataset and to explore the capability of the unsupervised models, we also manually prepared other datasets.

The new ProofWiki --- Mizar datasets were prepared by collecting sentences from MML and \textit{proofwiki.org}.
ProofWiki is a website that contains more than 26K theorems with detailed human-written proofs.
These proofs have been collectively contributed by many users over the years and are written in MathJax, a browser-adapted LaTeX format.
During 2015--2016, Grzegorz Bancerek embarked on an individual project to align ProofWiki theorems with Mizar theorems.
Bancerek manually aligned 470 theorems, more specifically theorem statements that appear in both MML and ProofWiki.
All the aligned theorems were in point-set topology and lattice theory.

Thanks to this effort we
are 
able to prepare a dataset similar to the English-to-French dataset used in UNMT.
This ProofWiki --- Mizar dataset contains two monolingual corpora, one in the Mizar language and the other in ProofWiki's LaTeX.
These two monolingual corpora need not be aligned.
The ProofWiki --- Mizar dataset also contains a smaller aligned corpus consisting of theorem statements extracted from Bancerek's manual alignment.
Since Bancerek's alignment only focuses on topology and lattice theory, we also picked sentences from the monolingual corpora that are from topology and lattice theory and form a dataset with the same aligned corpus but two smaller monolingual corpora.
Some statistics of these datasets are presented in Table~\ref{t:pwmiz} and sentence examples are given in Table~\ref{t:pwmizex}.

\begin{table}[h!]
\renewcommand{\arraystretch}{1.2}
\caption{The ProofWiki --- Mizar datasets}\label{t:pwmiz}
\begin{tabular}{m{0.74\columnwidth}<{\raggedright\arraybackslash}m{0.16\columnwidth}<{\raggedleft\arraybackslash}}
\toprule
\multicolumn{2}{c}{ProofWiki --- Mizar full dataset} \\
\midrule
Total sentences in monolingual Mizar & 1056478 \\
Unique sentences in monolingual Mizar & 1035511 \\
Total sentences in monolingual ProofWiki & 198801 \\
Unique sentences in monolingual ProofWiki & 198225 \\
Total sentences in pairs & 330 \\
\midrule
& \\
\midrule
\multicolumn{2}{c}{ProofWiki --- Mizar topology dataset} \\
\midrule
Total sentences in monolingual Mizar & 49408 \\
Unique sentences in monolingual Mizar & 48774 \\
Total sentences in monolingual ProofWiki & 29763 \\
Unique sentences in monolingual ProofWiki & 29618 \\
Sentences in pairs & 330 \\
\bottomrule
\end{tabular}
\end{table}

\begin{table}[h!]
\renewcommand{\arraystretch}{1.2}
\caption{The Mizar --- TPTP dataset}\label{t:miztptpstats}
\begin{tabular}{m{0.65\columnwidth}<{\raggedright\arraybackslash}m{0.25\columnwidth}<{\raggedleft\arraybackslash}}
\toprule
Total sentence pairs & 53994 \\
Unique sentence pairs & 53994 \\
Shortest Mizar sentence & 3 tokens \\
Shortest TPTP sentence & 2 tokens \\
Longest Mizar sentence & 986 tokens \\
Longest TPTP sentence & 786 tokens \\
\bottomrule
\end{tabular}
\end{table}

\begin{table*}[!htb]
\renewcommand{\arraystretch}{1.2}
\caption{Sample aligned ProofWiki --- Mizar theorem statements}\label{t:pwmizex}
\begin{tabular}{m{0.05\linewidth}<{\raggedright\arraybackslash}m{0.9\linewidth}<{\raggedright\arraybackslash}}
\toprule
Mizar & \texttt{\small for T being non empty TopSpace for A being Subset of T st A is countable holds A \string^0 = \{ \} } \\
LaTeX & Let T = \textbackslash left ( \{ S , \textbackslash tau \} \textbackslash right ) be a topological space . Let A be a subset of S . Then if A is countable , then A \string^ 0 = \textbackslash varnothing . \\
\cmidrule{2-2}
Mizar & \texttt{\small for T being non empty TopSpace for A , B being Subset of T holds ( A \textbackslash/ B ) \string^0 = ( A \string^0 ) \textbackslash/ ( B \string^0 ) } \\
LaTeX & Let T = \textbackslash left ( \{ S , \textbackslash tau \} \textbackslash right ) be a topological space . Let A , B be subsets of S . Then \textbackslash left ( \{ A \textbackslash cup B \} \textbackslash right ) \string^ 0 = A \string^ 0 \textbackslash cup B \string^ 0 \\
\bottomrule
\end{tabular}
\end{table*}

\subsection{The Mizar --- TPTP Dataset}\label{ss:miz-tptp}

As mentioned earlier, a serious disadvantage of using the Mizar language as our formal language is that we lose the ability to further improve the translation quality by harnessing Mizar's own toolchain.
Adding to the 
fact that the Mizar engine is closed-source, Mizar texts are notoriously hard to compile without a proper environment declaration, let alone a single Mizar statement without any external reference.
A correct environment declaration requires a conceptual understanding of the internals of the Mizar engine together with a lot of domain expertise.
This would be too challenging for us at the current stage if we want to automate the lookup of the environment declaratives.

To bypass this difficulty while fully exploring the capability of neural machine translation models, we design a different experiment in which the supervised NMT model takes Mizar statements as input and outputs TPTP formulas.
These TPTP formulas, after preprocessing, can be fed into a custom type elaboration tool to determine the correctness of a Mizar statement.
The aligned Mizar --- TPTP dataset can be generated with the MPTP toolchain~\cite{mptp0.2}. The size of the dataset is presented in Table~\ref{t:miztptpstats}.

\begin{table*}[!htb]
\renewcommand{\arraystretch}{1.2}
\caption{Sample Mizar --- TPTP statements}\label{t:miz-tptp}
\begin{tabular}{m{0.1\linewidth}<{\raggedright\arraybackslash}m{0.8\linewidth}<{\raggedright\arraybackslash}}
\toprule
Mizar (Source) & \texttt{\small for A holds A is doubleLoopStr  \& not A is empty implies for B holds B is Scalar of A implies B is being\_a\_square iff ex C st C is Scalar of A \& B = C \string^2} \\
\cmidrule{2-2}
Prefix (Target) & \texttt{\small c! b0  c=>\_\_2 c\&\_\_2 cnl6\_algstr\_0\_\_1 b0 c\~{}\_\_1 cnv2\_struct\_0\_\_1 b0  c! b1  c=>\_\_2 cnm4\_vectsp\_1\_\_2 b1 b0 c<=>\_\_2 cnv1\_o\_ring\_1\_\_1 b1  c? b2  c\&\_\_2 cnm4\_vectsp\_1\_\_2 b2 b0 cnr1\_hidden\_\_2 b1 cnk1\_o\_ring\_1\_\_1 b2} \\
\cmidrule{2-2}
TPTP-FOF & \texttt{\small fof(a1,elaborate,(! [A0] : ((nl6\_algstr\_0(A0) \& (\~{} nv2\_struct\_0(A0))) => (! [A1] : (nm4\_vectsp\_1(A1,A0) => (nv1\_o\_ring\_1(A1) <=> (? [A2] : (nm4\_vectsp\_1(A2,A0) \& nr1\_hidden(A1,nk1\_o\_ring\_1(A2)))))))))).} \\
\cmidrule{2-2}
TPTP-THF & \texttt{\small thf(a1,elaborate, (! [A:\$i]:(((nl6\_algstr\_0 @ A) \& \~{} ((nv2\_struct\_0 @ A))) => (! [B:\$i] : ((nm4\_vectsp\_1 @ B @ A) =>  ((nv1\_o\_ring\_1 @ B) <=>  (? [C:\$i] : ((nm4\_vectsp\_1 @ C @ A) \& (nr1\_hidden @ B @ (nk1\_o\_ring\_1 @ C)))))))))).} \\
\cmidrule{2-2}
Elaborated-FOF & \texttt{\small fof(a1,conjecture,(! [X0] : ((\~{}(l6\_algstr\_0(X0) => (\~{}(\~{}v2\_struct\_0(X0))))) => (! [X1] : (m1\_subset\_1(X1,u1\_struct\_0(X0)) => (v1\_o\_ring\_1(X1,X0) <=> (\~{}(! [X2] : (\~{}(\~{}(m1\_subset\_1(X2,u1\_struct\_0(X0)) => (\~{}r1\_hidden(X1,k1\_o\_ring\_1(X0,X2)))))))))))))).} \\
\bottomrule
\end{tabular}
\end{table*}

\section{Neural Machine Translation Models}\label{s:nn}

\subsection{Brief Overview}
Theoretical foundations demonstrating the expressiveness of neural networks date back to late 1980s.
In 1989, Cybenko~\cite{Cybenko1989} and Hornik at al.~\cite{Hornik1989} showed that any measurable function between Euclidean spaces can be uniformly approximated by a composition of an affine mapping together with a non-linear mapping satisfying certain mild restrictions.
Hornik (1991)~\cite{Hornik1991} further generalized the underlying spaces and relaxed the restrictions on the non-linear mapping.
Due to the limitation of computing resources as well as the scarcity of large datasets, we did not see competitive applications based on neural network until the seminal work of AlexNet in 2012~\cite{imagenet} which harnessed the computational power of GPU and the ImageNet dataset.
Since then, machine learning based on deep neural networks has gained tremendous popularity in a wide range of areas from computer vision~\cite{imagenet} to natural language translation~\cite{chogru} and even to mastering the game of Go~\cite{alphago}.
During these years we have also witnessed progress in learning theory:
Telgarsky (2015)~\cite{Telgarsky2015} and Safran et al. (2016)~\cite{Safran2016} showed that the increase of depth in a deep neural network can exponentially reduce the width of the affine maps at each layer and
Du (2018)~\cite{Du2018} discovered the global convergence of gradient descent in an over-parameterized neural network.

In neural machine translation, a series of works published around 2014 has laid out the foundation for architecture based on recurrent neural networks (RNNs):
Chung at al. first extended RNN with sophisticated recurrent units that implement a gating mechanism~\cite{chogru};
Sutskever at al.~\cite{sutskever-seq2seq-lstm} incorporated the Long Short-Term Memory cell (LSTM)~\cite{lstm} into RNN and discovered the importance of bi-directional encoding;
Bahdanau et al.~\cite{bahdanau-attention} and Luong at al.~\cite{LuongPM15} discovered the attention mechanism and demonstrated its effectiveness in handling long sequences.
These contributions eventually led to Luong's NMT implementation~\cite{luong17}.

Meanwhile architectures that are not based on RNN have also been developed:
Gehring et al. (2017)~\cite{gehring2017convs2s} applied convolutional neural networks (CNNs) in sequence to sequence translation.
Vaswani et al. (2017)~\cite{vaswani2017} invented a self-attention mechanism called the Transformer that overcomes the lack of parallelization in RNN-based training.
This parallelization of encoding also enabled the use of pre-training which led to the seminal BERT model developed by Devlin et al.~\cite{bert}
The UNMT model~\cite{lample2018phrase} we are using is based on the Transformer architecture, adding a novel back translation technique for unsupervised learning.
The XLM model~\cite{lample2019cross} extends the UNMT model with BERT-style pretraining.
Unsupervised learning of UNMT is then performed on the pre-trained model to fine-tune the model parameters.

\subsection{RNN-Based NMT}
The fundamental idea behind neural machine translation is to encode source and target sentences as lists of numeral vectors and feed them into a network architecture for computation of a loss function.
Since the neural network is essentially a differentiable function, training the neural network amounts to minimization of the loss function.
This is a gradient optimization problem and in neural networks the gradients of network parameters are computed by the back-propagation algorithm.

At the top level, the NMT architecture consists of an encoder network and a decoder network, each of which is a multi-layer recurrent neural network with LSTM cell as its network cells.

During training, the encoder takes in a source sentence one word vector at a time.
After all the word vectors have been put in, an end-of-sentence marker initiates the decoder which takes in the hidden vector of the encoder.
The decoder produces a target sentence one word vector at a time and simultaneously takes in the correct target sentence.
The evaluated target sentence is then compared with the correct target sentence and the loss value is computed.

During inference, only a source sentence is needed.
The source sentence is first taken by the encoder.
The encoder then passes the hidden vector to the decoder.
The decoder then produces the evaluated target sentence one word vector at a time.
Each time a word vector is generated it is put directly back to the decoder input for the evaluation of the next word vector.
There are two ways to pick a word vector. First, given a probability distribution of words, the \emph{greedy} approach always picks the next word as the one with the highest probability. Second,
the \emph{beam search} approach also picks the next word from the probability distribution of words, but it is chosen in such a way the sentences with the highest $n$ joint probabilities are kept ($n$ is the \textit{width} of the beam search).
We use the greedy approach in our model comparison experiments for the fairness of the comparison, while we will use the beam search approach to further augment the data in Section \ref{s:elab}.

One crucial add-on to the encoder-decoder architecture is the attention mechanism, which can be considered as a side network connecting the encoder and the decoder, forming a larger network.
The attention network records the hidden vector at each word input of the source sentence.
Depending on the type of attention, different computation steps are involved to obtain an attention vector.
This attention vector is then fed into the decoder to affect the generation of the target sentence.

Our NMT experiment uses 2-layer LSTM cells in both the encoder and the decoder.
The dimension of hidden vectors in a LSTM cell is 1024.
We used scaled Luong's attention, bi-directional encoding and 20\% dropout rate.
Due to the size of this network we cannot fit into a single GPU, so we train our model in CPU-only mode.
The total training time is 17.9 hours.
The synthetic dataset is divided into a 90:10 ratio between training and inference set.
An extra 4000 sentence pairs are reserved, in which 2000 are used for model validation and the remaining half are used for testing and model evaluation.

\subsection{UNMT}

\begin{figure}[h!]
  \includegraphics[scale=0.35]{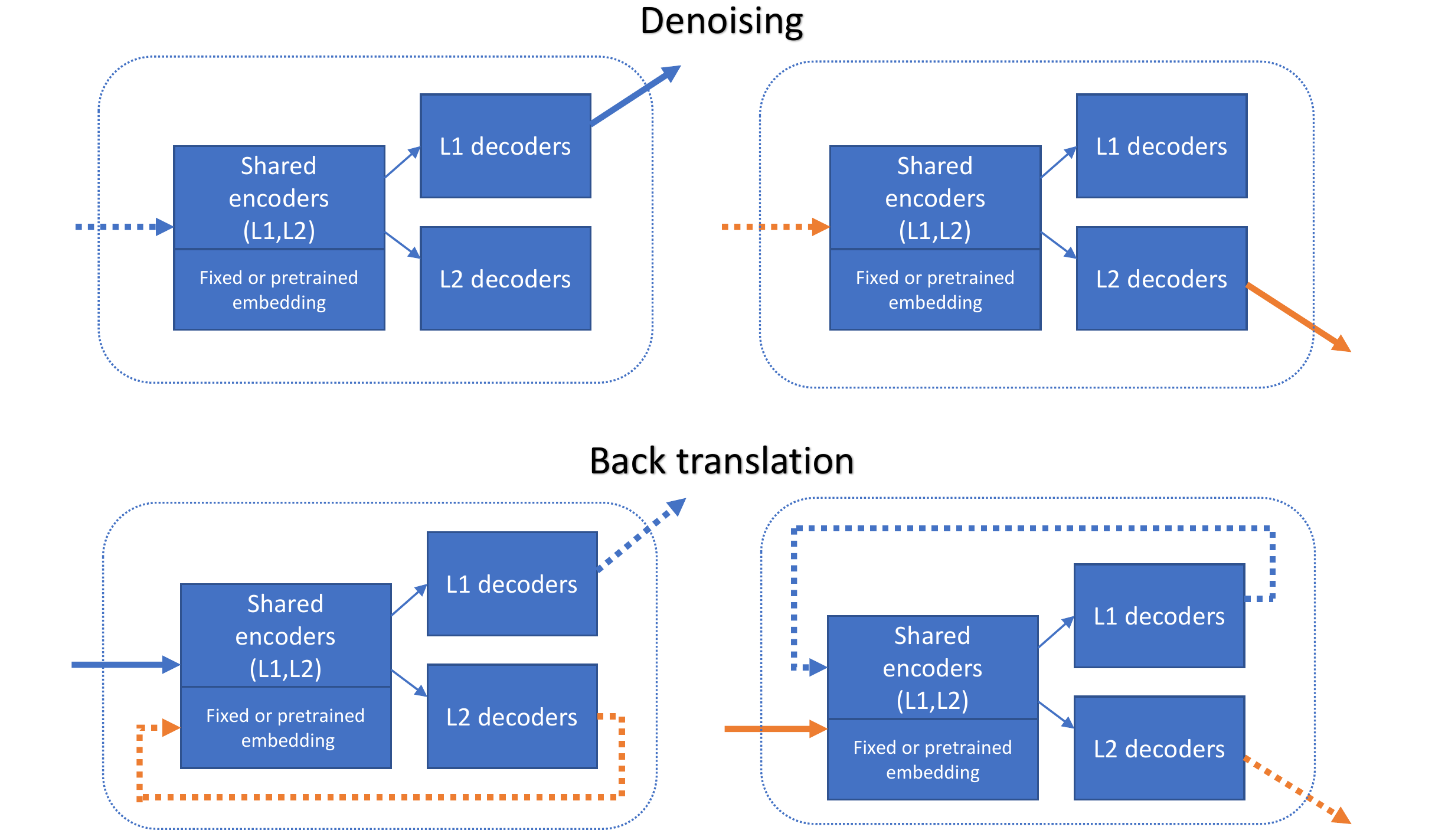}
  \caption{Denoising and back translation in UNMT and XLM}
  \label{fig:unsup}
\end{figure}

The key idea of the unsupervised learning according to Lample et al.~\cite{lample2018phrase} is to transform the unsupervised learning problem into a series of supervised learning problems.
This idea is illustrated in Figure~\ref{fig:unsup}.
First, the two monolingual corpora are concatenated and are processed by fastText~\cite{fasttext}, a word2vec~\cite{word2vec} implementation for learning word representation.
This bilingual word representation is kept fixed throughout the training and evaluation phase.

The neural network architecture of UNMT consists of a shared encoder and two decoders.
To initialize the parameter values, a denoising step is performed: for each monolingual corpus, we corrupt each sentence by randomly permuting and dropping a few words.
We then use the corrupted corpus as source language and the original non-corrupted corpus as target language.
This gives us two aligned corpora.
We use one of them for initializing the parameters of the shared encoder and the L1 decoder.
And the other for refining the parameters of the shared encoder and initializing the parameters of the L2 decoder.

The training is done using a back translation technique.
For each sentence in language L1 as input to the neural network, we first generate a sentence in L2 from the L2 decoder.
This generated sentence is then put back into the shared encoder again and we now generate a sentence in L1 from the L2 decoder.
The combined result is a merged network taking L1 sentence as input and generating L1 sentence as output, and this can be treated as a supervised problem.
Similarly we can do the same on L2, generating L2 $\rightarrow$ L1 $\rightarrow$ L2 data flow and obtain an L2 $\rightarrow$ L2 supervised problem.
The UNMT model is trained in alternative fashion: a few iterations L1 $\rightarrow$ L1 followed by a few iterations of L2 $\rightarrow$ L2.
This alternation continues until a reasonable stopping criterion is met.

In UNMT we can configure the encoder-decoder architecture as either based on RNN or based on Transformer.
In our experiments we use the Transformer architecture, since it provides a similar quality, while allowing for more efficient training on GPU architectures.
The key feature in a Transformer block is a multi-headed attention cell, which is a multi-copy version of so called self-attention cells.
A self-attention cell computes several feature vectors for each word in a sentence in parallel.
Then a cross-weighting and averaging step is computed.
Since the weighting step is computed in parallel and the input sentence is read in parallel, the Transformer cell can maximally harness the power of GPU and is usually trained much faster than RNN.

The UNMT model was trained on one NVIDIA GeForce GTX 1080 Ti Graphics Card. The training times are 30 minutes for the ProofWiki --- Mizar topology dataset and 1 hour 8 minutes for the ProofWiki --- Mizar full dataset.

\subsection{XLM}

The key difference between UNMT and XLM is that in UNMT the word embedding is fixed throughout the training process while in XLM it is also pre-trained.
XLM uses Masked Language Model (MLM) method borrowed from BERT~\cite{bert}.
In MLM, each word in a sentence has 15\% chance of being selected.
If a word is selected, it has 80\% chance of being substituted by a \texttt{[MASK]} token; 10\% chance of being substituted by a random token; and 10\% chance being unchanged.
After pre-training, back translation similar to UNMT is performed to fine-tune the parameters in both encoders and decoders.

The XLM model was also trained on one NVIDIA GeForce GTX 1080 Ti Graphics Card. The training times are 4 hours 24 minutes for the ProofWiki --- Mizar topology dataset and 7 hour 53 minutes for the ProofWiki --- Mizar full dataset.

\section{Experiments and Results}\label{s:exp}

We compare the three models introduced in Section~\ref{s:nn} by three metrics that are commonly used in evaluating the quality of language translation models: the BLEU rate~\cite{bleu}, the perplexity score and the Levenshtein edit distance.

The calculation of the BLEU rate involves counting of $n$-grams, geometric averaging, as well as an exponentially scaled length penalty.
It computes a score from 0 to 100, with larger values indicating better translation quality.
In NLP research, bilingual machine translation attains a typical BLEU score ranging from 25.16 to 40.51 depending on the languages (see Table 1. of~\cite{gehring2017convs2s}).

The perplexity also involves counting of $n$-grams with exponentials.
It intuitively captures the idea of the number of words randomly picked before obtaining the correct word.
It is a positive number with smaller value denoting better translation quality.
In Table 4 and 5 of~\cite{gehring2017convs2s} we witnessed the perplexities ranging between 6 to 7 in bilingual translation.

The edit distance is also a very common evaluation metric. It calculates the number of insertions, deletions and substitutions from a translated sentence to its corresponding correct sentence.
The calculation is done using a dynamic programming algorithm.
The smaller the edit distance is, the closer the translated sentence to correctness.

It is noted that all the above three metrics only compare the syntactic closeness of a translated corpus from the reference corpus.
Although in terms of formalization evaluating syntactic closeness is primitive (e.g. missing a `not' in the statement negates its meaning), for convenience and speed we temporarily borrow what is available in existing NLP research.
It would be interesting if new techniques on measuring closeness of formal statements by semantics could appear.

Table~\ref{t:eval-synthetic} gives the metrics of a 2000-sample evaluation on the synthetic dataset.
Because of the maximum use of alignment, we can see that NMT performs consistently better than both UNMT and XLM in all metrics.
As XLM uses a pre-training step instead of fixing a word embedding, we also anticipate that XLM performs better than UNMT.
This is indeed verified by the results shown.

For truly non-aligned datasets, we can only compare the two unsupervised models (Table~\ref{t:eval-pfwk}).
Due to the small size of our aligned corpus (as well as the fact that each sentence pair is generally longer than the synthetic dataset), we witness a significant decrease of performance in all the metrics.
However, if we compare within the dataset, we can still see a trend that, except the surprising perplexity score, 
XLM performs generally better than UNMT.
In addition, since our aligned corpus is only focused on topology and lattice theory, we see the trend that no matter which model we used, the metrics show better results on the topology sub-dataset instead of the full ProofWiki --- Mizar dataset.

\begin{table}[!ht]
\renewcommand{\arraystretch}{1.2}
\caption{Model evaluation on the synthetic dataset}\label{t:eval-synthetic}
\begin{tabular}{m{0.3\columnwidth}>{\centering\arraybackslash}m{0.17\columnwidth}>{\centering\arraybackslash}m{0.17\columnwidth}>{\centering\arraybackslash}m{0.17\columnwidth}}
\toprule
{\small 2000 samples} & NMT & UNMT & XLM \\
\midrule
BLEU & 70.9 & 27.14 & 43.61 \\
Perplexity & 1.58 & 3.00 & 2.91 \\
Edit distance 0 & 65.2\% & 26.8\% & 34.1\% \\
Edit distance $\leq$1 & 74.6\% & 34.4\% & 38.5\% \\
Edit distance $\leq$2 & 81.5\% & 41.8\% & 42.1\% \\
Edit distance $\leq$3 & 83.9\% & 46.3\% & 45.9\% \\
\bottomrule
\end{tabular}
\end{table}

\begin{table}[!ht]
\renewcommand{\arraystretch}{1.2}
\caption{Model evaluation on the Proofwiki --- Mizar dataset}\label{t:eval-pfwk}
\begin{tabular}{m{0.3\columnwidth}>{\centering\arraybackslash}m{0.1275\columnwidth}>{\centering\arraybackslash}m{0.1275\columnwidth}>{\centering\arraybackslash}m{0.1275\columnwidth}>{\centering\arraybackslash}m{0.1275\columnwidth}}
\toprule
{\small 131 samples} & \multicolumn{2}{c}{UNMT} & \multicolumn{2}{c}{XLM} \\
\cmidrule(lr){2-3} \cmidrule(lr){4-5}
& topology & full & topology & full \\
\midrule
BLEU & 4.03 & 1.55 & 7.87 & 6.07 \\
Perplexity & 11.57 & 10.73 & 33.01 & 39.70 \\
Edit distance 0 & 0\% & 0\% & 0\% & 0\% \\
Edit distance $\leq$1 & 0\% & 0\% & 0\% & 0\% \\
Edit distance $\leq$2 & 0\% & 0\% & 0.76\% & 0\% \\
Edit distance $\leq$3 & 9.92\% & 2.29\% & 6.11\% & 2.29\% \\
\bottomrule
\end{tabular}
\end{table}

\section{Data Augmentation Using Type Elaboration}\label{s:elab}

\newcommand{\patt}[1]{{\underline{#1}}}
\newcommand{\pred}[1]{\mathsf{pred}^{#1}}
\newcommand{\prim}[1]{\mathsf{prim}^{#1}}
\newcommand{\limplies}{\to}
\newcommand{\lequiv}{\equiv}
\newcommand\reals{{\mathsf{R}}}
\newcommand\realp{{\mathsf{real}}}
\newcommand\natp{{\mathsf{nat}}}
\newcommand\elaborator{{\sc{Elaborator}}}

The major motivation of this experiment is to form a feedback loop such that the evaluated target sentence can be deterministically transformed back to a source sentence.
By using beam search during the inference phase of the NMT model, we can generate more than one target sentence.
If some of these target sentences can be translated back to the source sentences, we can add the new sentence pairs into our training dataset and retrain our model.
We anticipate that the model with augmented data will have more syntactically correct translation, or in a way, we manage to ``teach'' the model the correct syntactic format.

In Section~\ref{ss:elab-intro} we shortly recall the soft type system in order to cover the internal mechanism of our {\elaborator} that does Mizar-style type checking in Section~\ref{ss:elab-type}.
In Section~\ref{ss:elab-pattern} we illustrate how pseudo-patterns in a TPTP-THF formula are \textit{type-checked} by the {\elaborator}.
We integrate the {\elaborator} and our data transformation pipeline into the feedback loop in Section~\ref{ss:elab-pipeline} and show experimental results in Section~\ref{ss:elab-exp}.

\subsection{Mizar Soft Type System}\label{ss:elab-intro}

All Mizar types are represented as predicates (soft types) in MPTP~\cite{mptp0.2}.
Hence even a Mizar type corresponding to the real numbers would be
represented by a predicate over a type of individuals (sets).

Suppose we are given a binary operation $+$ (written in infix), a constant ${\hat{1}}$
and a unary predicate $\realp$.
For each positive natural number $n$, let $\hat{n}$ be the term
${\hat{1}}+\cdots +{\hat{1}}$ (associating to the left).
Consider the following three formulas.
\begin{enumerate}
\item $\realp~{\hat{1}}$
\item $\forall x y.\realp~x~\to~\realp~y~\to~\realp~(x+y)$
\item $\forall x y.\realp~x~\to~\realp~y~\to~x+y = y+x$
\end{enumerate}
In order to prove the conjecture
$\forall x.\realp~x~\to~x + {\hat{n}} = {\hat{n}} + x$
a theorem prover would need to 
instantiate using the commutativity axiom and then prove $\realp~{\hat{n}}$.
Proving $\realp~{\hat{n}}$ requires applying the second axiom $n-1$ times.

The formula above indicating the sum of two reals is real
is an example of a ``function type'' axiom,
indicating that if the inputs of a function has certain soft types
then the result will have certain soft types.
A different kind of soft typing axiom is a ``type hierarchy'' axiom,
indicating that if a term has certain soft types
then it also has certain ``inherited'' soft types.
We can extend the example to include a type hierarchy axiom by adding
a predicate for natural numbers.

Suppose we add a unary predicate $\natp$
and change the axioms of the problem to be the following:
\begin{enumerate}
\item (Function type for ${\hat{1}}$) $\natp~{\hat{1}}$
\item ($\realp$ function type for $+$) \\$\forall x y.\realp~x~\to~\realp~y~\to~\realp~(x+y)$
\item ($\natp$ function type for $+$) \\$\forall x y.\natp~x~\to~\natp~y~\to~\natp~(x+y)$
\item (Type hierarchy for $\natp$) $\forall x.\natp~x~\to~\realp~x$
\item $\forall x y.\realp~x~\to~\realp~y~\to~x+y = y+x$  
\end{enumerate}
The conjecture $\forall x.\realp~x~\to~x + {\hat{n}} = {\hat{n}} + x$
still follows. In this case, however, a proof
must make use of the type hierarchy axiom to infer that ${\hat{n}}$ is real.
There are actually multiple derivations.
At one extreme one can infer that ${\hat{1}}$ is real and
use the $\realp$ function typing axiom for $+$ $n-1$ times.
At the other extreme one can use the $\natp$ function typing axiom for $+$ $n-1$ times
to infer that $\hat{n}$ is a natural number and then infer that $\hat{n}$ is real.
For a general theorem proving procedure this redundancy leads to a needless
expansion of the search space.
By treating soft types as special, the fact that $\hat{n}$ is real will
be computed by the (deterministic) soft typing algorithm.

Until now we have only considered the unary predicates $\realp$ and $\natp$
as soft types.
In order to handle predicates corresponding to Mizar soft types
we must be able to handle predicates with extra dependencies.
For example, in Mizar there are types ``Element of $X$''
and ``Function of $A$, $B$.''
These could appear as a binary predicate ${\mathsf{elt}}$
and a ternary predicate ${\mathsf{func}}$ in a problem,
where we write ${\mathsf{elt}}~x~X$ to mean $x$ has type ``Element of $X$''
and write ${\mathsf{func}}~f~A~B$ to mean $f$ has type ``Function of $A$, $B$.''
We have also considered variants of our example with soft typing axioms such as
\begin{itemize}
\item (Function type for ${\mathsf{ap}}$) $\forall a b f.{\mathsf{func}}~f~a~b\to\forall x.{\mathsf{elt}}~x~a\to{\mathsf{elt}}~({\mathsf{ap}}~a~b~f~x)~b$
\end{itemize}
where ${\mathsf{ap}}$ is a $4$-ary operator for set theoretic function application.

\subsection{Mizar Soft Type Inference}\label{ss:elab-type}

In~\cite{Urban06MoMM} the notion of a (first-order) ``Mizar-like Horn theory'' is defined,
distinguishing between a ``type hierarchy part'' and ``functor types part.''
We lift this to the higher-order setting by carving out appropriate first-order formulas
from higher-order formulas~\cite{Church40}
satisfying similar conditions.

\begin{itemize}
\item A {\emph{first-order term}} is either a variable (of base type)
  or a term of the form $f\,t_1\,\ldots\,t_n$ where $f$ is an $n$-ary function (taking $n$ terms of base type to a term of base type)
  and $t_i$ is a first-order term for each $i\in\{1,\ldots,n\}$.
\item A {\emph{first-order atom}} is $p\,t_1\,\cdots\,t_n$
  where $n\geq 1$, $p$ is an $n$-ary predicate over terms of base type
  and $t_i$ is a first-order term for each $i\in\{1,\ldots,n\}$.
  Given a first-order atom $A$ of the form $p\,t_1\,\cdots\,t_n$,
  we say $p$ is the predicate of $A$ and write $\pred{A} = p$
  and say $t_1$ is the primary term of $A$ and write $\prim{A} = t_1$.
\item A {\emph{first-order literal}} is either a first-order atom or the negation of a first-order atom.
  We extend $\pred{L}$ and $\prim{L}$ to be $\pred{\neg A} = \pred{A}$ and $\prim{\neg A} = \prim{A}$
  and call $\pred{L}$ the predicate of $L$ and
  $\prim{L}$ the primary term of $L$.
  We say $L$ has positive polarity if it is an atom and negative polarity otherwise.
\end{itemize}

In~\cite{Urban06MoMM} a chronological total order $<$ on functor and type symbols is used
to add appropriate conditions on Mizar-like Horn clauses.
Here we instead work with an strict partial order $\prec$ which is computed dynamically
while determining the classification of axioms.

Let $\prec$ be a strict partial order on constants.
Let $\varphi$ be a formula
$$\forall x_1 \cdots \forall x_n . L_1 \land\cdots\land L_m \limplies L_{m+1} \land\cdots\land L_{m+k}$$
\begin{itemize}
\item We say $\varphi$ is a {\emph{type hierarchy formula}} if
  \begin{enumerate}
  \item $n\geq 1$,
  \item $m > 1$,
  \item $L_m$ is either of the form $p x_1\ldots x_n$ or $\neg p x_1\ldots x_n$,
  \item for each $i\in\{1,\ldots,m-1\}$
    $\prim{L_i}\in\{x_1,\ldots,x_{n-1}\}$
    and $x_n$ is not free in $L_i$
    and
  \item for each $i\in\{m+1,\ldots,m+k\}$
    we have $\prim{L_i} = x_n$
    and $\pred{L_i} \prec p$.
  \end{enumerate}
\item We say $\varphi$ is a {\emph{function type formula \textup{(}for $f$\textup{)}}} if for each $i\in\{1,\ldots,m\}$
  \begin{enumerate}
  \item $\prim{L_i}\in\{x_1,\ldots,x_n\}$ and
  \item for each $j\in\{m+1,\ldots,m+k\}$
    $\prim{L_j}$ is $f\,x_1\cdots x_n$
    and $\pred{L_i} \prec f$.
  \end{enumerate}
\end{itemize}

We now describe the soft typing inference mechanism designed for
a type system given by function type and type hierarchy formulas.
We first describe a procedure for ``widening'' a given soft type
and then describe how all soft types of $t$ are computed.

Suppose we have computed a soft typing literal $L$.
We can compute new soft types by a ``widening'' procedure making use of type hierarchy formulas
as follows.
\begin{itemize}
\item Find all type hierarchy formulas
  $$\forall x_1 \cdots \forall x_n . L_1 \land\cdots\land L_m \limplies L_{m+1} \land\cdots\land L_{m+k}$$
  where $L_m$ has the same predicate and polarity as $L$.
  Let $\theta$ be a substitution such that $\theta(L_m) = L$. (The restrictions on type hierarchy formulas
  ensure such a $\theta$ exists.)
  Continue if $\theta(L_j)$ has already been computed to be a soft type of its principle term for each $j\in\{1,\ldots,m-1\}$.\footnote{Note that this may give different results depending on the order in which soft types have been computed.}
\item For each $j\in\{m+1,\ldots,m+k\}$, $\theta(L_j)$ is a new soft type for the same term as the primary term of $L$.
  We add this as a known soft type and recursively widen each such $\theta(L_j)$.\footnote{Note that the $\prec$ relation guarantees termination.}
\end{itemize}
  
We now describe the algorithm for computing all soft types of
a ground first-order term $t$.
The algorithm proceeds by recursion.
\begin{enumerate}
\item Since $t$ is ground, it must have the form $f\,t_1\,\ldots\,t_n$
  where each $t_i$ is a ground first-order term.
  Assume all soft types of each $t_i$ have been computed.
\item We next find all function type formulas for $f$, e.g.,
  $$\forall x_1 \cdots \forall x_n . L_1 \land\cdots\land L_m \limplies L_{m+1} \land\cdots\land L_{m+k}$$
  where $f(x_1,\ldots,x_n)$ is the primary term of $L_{j}$ for $j\in\{m+1,\ldots,m+k\}$.
  Let $\theta$ be the substitution with $\theta(x_i) = t_i$ for $i\in\{1,\ldots,n\}$.
\begin{enumerate}
\item Note that for each $j\in\{1,\ldots,m\}$
  the principle term of $\theta(L_j)$ is some $t_i$ and we have already
  computed the soft types of $t_i$.
  Continue if $\theta(L_j)$ is a known soft typing literal for each $j\in\{1,\ldots,m\}$.
\item For each $j\in\{m+1,\ldots,m+k\}$, add $\theta(L_{j})$ as a known soft typing literal
  and call the widening procedure with $\theta(L_{j})$.
\end{enumerate}
\end{enumerate}

The procedure described above is a simplification of the
one used by the {\elaborator}, as there are still more soft typing
formulas than function type and type hierarchy formulas.
For example, there are redefinition formulas allowing
one to declare that a new function is defined to be the
same as some previous function but with new typing information
for the new function. The equations introduced by redefinitions
require introducing some degree of equational reasoning
within the soft typing inference procedures.
For brevity we omit details.

\subsection{{\elaborator} --- Patterns and Elaboration}\label{ss:elab-pattern}

In order to allow for partially specified terms and formulas,
we extend the language to allow {\emph{pattern names}}.
A pattern name is used as a predicate or function with a fixed arity,
but should not be thought of as having a clear intended semantics.
Instead each pattern name will be associated with a different
{\emph{constructor name}} which can be thought of semantically.
We call terms and formulas with occurrences of pattern names
{\emph{pseudo-terms}} and {\emph{pseudo-formulas}}.
Our primary use of {\elaborator} is to try to compute
formulas corresponding to pseudo-formulas produced by the translation mechanism.

Recall the constructor name $\mathsf{ap}$ with function type formula
$$\forall a b f.{\mathsf{func}}~f~a~b\to\forall x.{\mathsf{elt}}~x~a\to{\mathsf{elt}}~({\mathsf{ap}}~a~b~f~x)~b.$$
Here ${\mathsf{func}}$ and ${\mathsf{elt}}$ are also constructor names.
For each of these constructor names there will be corresponding pattern
names ${\patt{\mathsf{ap}}}$, ${\patt{\mathsf{func}}}$ and ${\patt{\mathsf{elt}}}$.
Along with the formulas defining the type system we give {\elaborator}
a pseudo-formula for each pattern name.
For the pattern names above, three pseudo-formulas similar to the following would be given.
\begin{enumerate}
  \item $\forall A B.{\patt{\mathsf{elt}}}~B~A~\lequiv~{\mathsf{elt}}~B~A$.
  \item $\forall A B f.{\patt{\mathsf{func}}}~f~A~B\lequiv~{\mathsf{func}}~f~A~B$.
  \item $\forall C D f c.{\mathsf{func}}~f~C~D~\land~{\mathsf{elt}}~c~C~\limplies~{\patt{\mathsf{ap}}}~f~c = {\mathsf{ap}}~C~D~f~c$.
\end{enumerate}
In each case there is an equation or equivalence indicating how to elaborate
a pattern name into a term or formula using corresponding constructor name,
possibly guarded by soft typing constraints on the variables.
In the case of ${\patt{\mathsf{ap}}}$
the soft typing constraints determine the extra two arguments required by ${\mathsf{ap}}$.

Given these pattern pseudo-formulas in addition to the typing formulas,
the algorithm used by the {\elaborator} to elaborate pseudo-formulas into formulas is relatively straightforward.
The essential idea is to recursively traverse pseudo-formulas and pseudo-terms
elaborating from the bottom up. At each step there will be type constraints
required by the pattern pseudo-formulas and the type checker is used to check if these
type constraints are satisfied. If the type constraints are satisfied, terms
for the missing arguments will be determined.
When traversing binders, introduce a fresh name to replace the variable and assume this
fresh name has all appropriate soft typing assumption.
For example, to elaborate $\forall x.\varphi(x) \to \psi(x)$, we extract
the parts of $\varphi$ that correspond to soft typing literals,
create a fresh name $a$, assume $a$ satisfies these soft typing literals,
and then continue by trying to elaborate $\varphi(a)\to\psi(a)$.
Similar remarks apply when attempting to elaborate $\exists x.\varphi(x)\land\psi(x)$.

\begin{example}\label{ex:elab}
  Suppose we wish to elaborate the following pseudo-formula:
  $$\forall A x.{\patt{\mathsf{elt}}}~x~A~\limplies~\exists f.{\patt{\mathsf{func}}}~f~A~A~\land~{\patt{\mathsf{ap}}}~f~x = x.$$
  The elaborated formula is
  $$\forall A x.{\mathsf{elt}}~x~A~\limplies~\exists f.{\mathsf{func}}~f~A~A~\land~{\mathsf{ap}}~A~A~f~x = x.$$
  This elaborated formula is determined as follows.
  We choose fresh names $a$ and $y$ for $A$ and $x$ and try to elaborate
  the antecedent of the implication: ${\patt{\mathsf{elt}}}~y~a$.
  This succeeds giving ${\mathsf{elt}}~y~a$.
  We recognize this as a typing literal and assume this as typing information about $y$.
  We then begin elaborating the succedent of the implication.
  We choose a fresh name $g$ for $f$ and attempt to elaborate the left side of the conjunction:
  ${\patt{\mathsf{func}}}~g~a~a$.
  This succeeds giving ${\mathsf{func}}~g~a~a$ which is then assumed as type information about $g$.
  We finally elaborate ${\patt{\mathsf{ap}}}~g~y = y$.
  To elaborate ${\patt{\mathsf{ap}}}~g~y$ we must find types of $g$ and $y$
  of the form ${\mathsf{func}}~g~Y~Z$ and ${\mathsf{elt}}~y~Y$.
  Using the typing assumptions above we know
  ${\mathsf{func}}~g~a~a$ and ${\mathsf{elt}}~y~a$.
  Hence we can elaborate ${\patt{\mathsf{ap}}}~g~y$ as ${\mathsf{ap}}~a~a~g~y$.
  When passing back through the quantifiers, the fresh names are replaced with the
  original bound variables.
\end{example}

\subsection{Data Augmentation Pipeline}\label{ss:elab-pipeline}

\begin{figure}[h!]
  \includegraphics[scale=0.35]{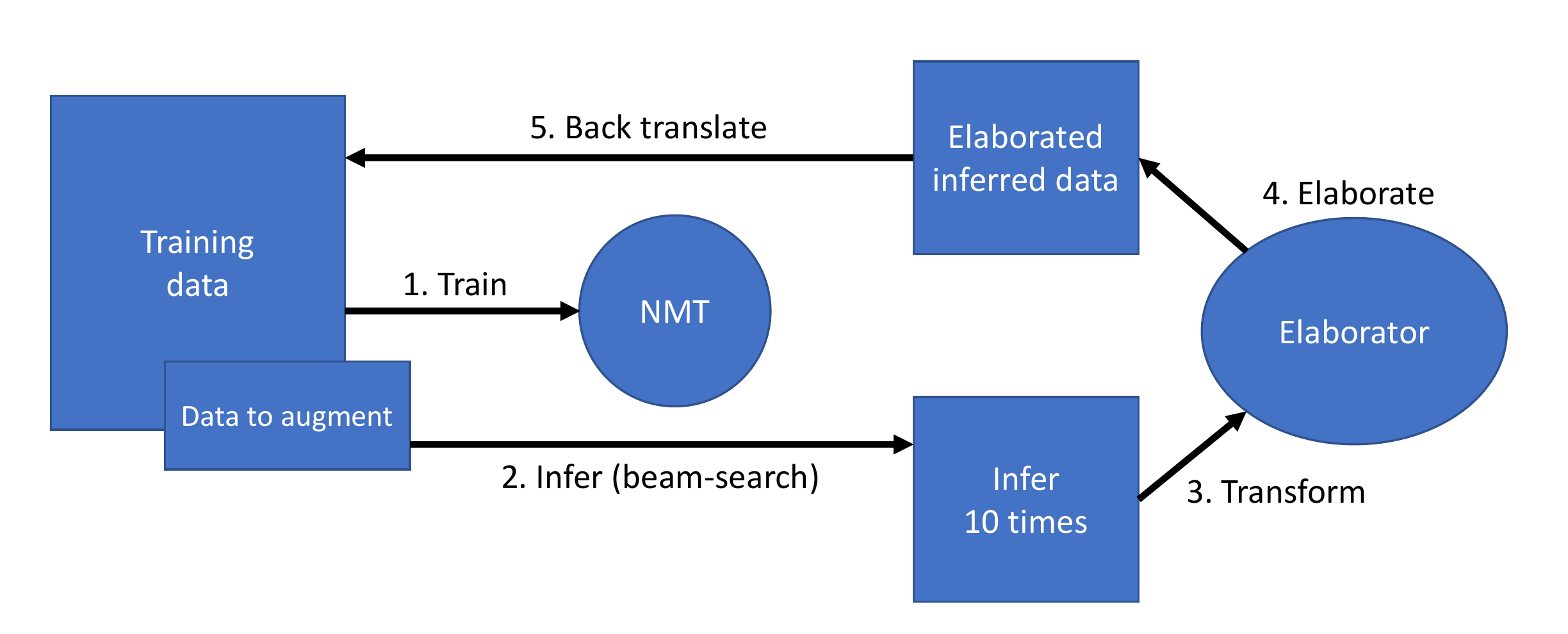}
  \caption{Feedback loop Supervised NMT + {\elaborator}}
  \label{fig:feedback-loop}
\end{figure}

\begin{figure}[h!]
  \includegraphics[scale=0.35]{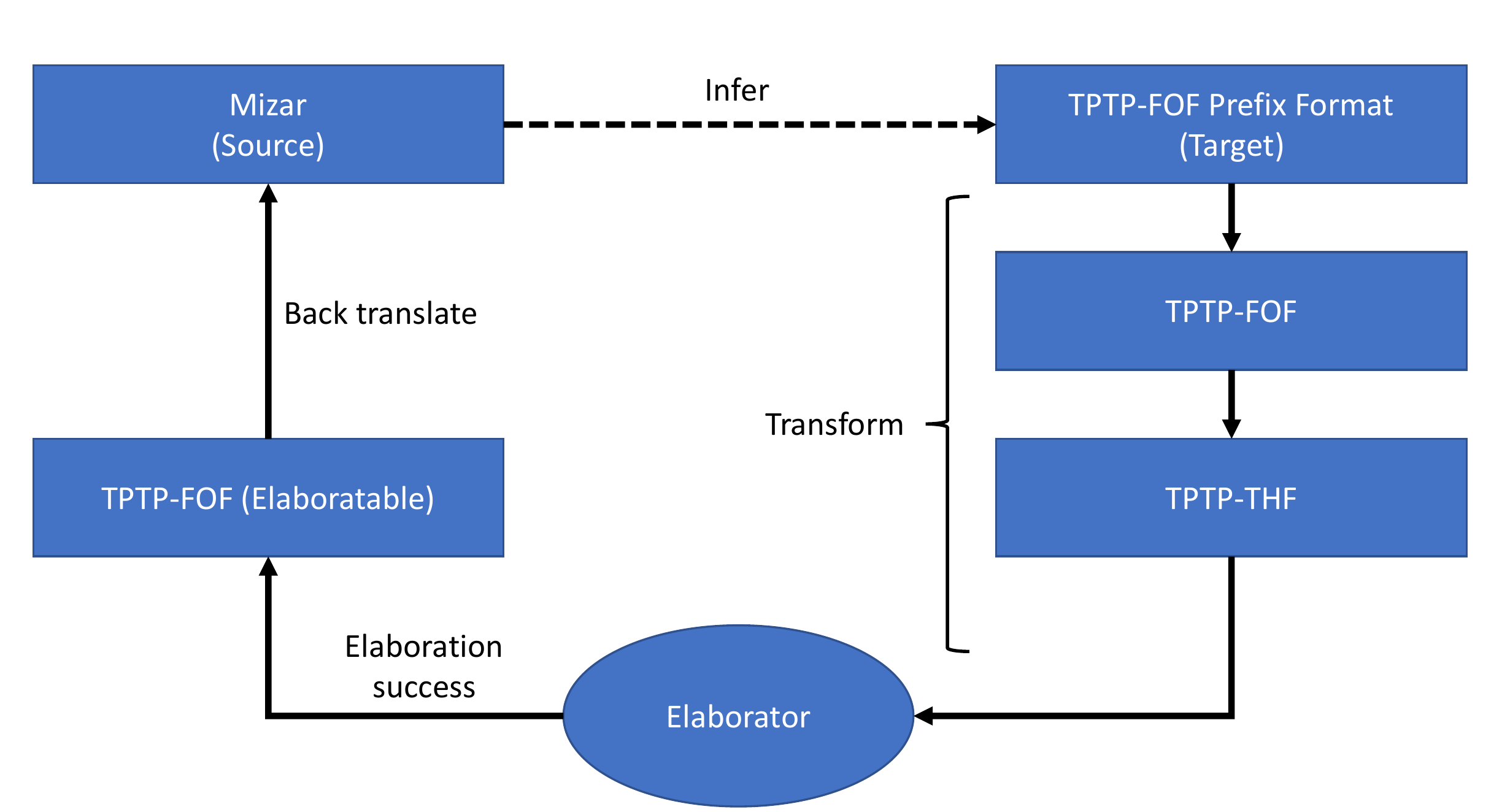}
  \caption{Format transformation pipeline}
  \label{fig:format}
\end{figure}

Figure~\ref{fig:feedback-loop} illustrates the idea of the feedback loop mentioned above.
In the chart the $x$-axis denotes edit distances while the $y$-axis denotes the number of sentences inferred.
We use the Mizar --- TPTP dataset as described in Section~\ref{ss:miz-tptp}.
Instead of using the standard TPTP first-order format, we use a prefix notation.
This notation not only largely reduces the need for matching parenthesis, but it also reduces the overall length of the TPTP sentence.
This has been previously found to benefit the neural translation task~\cite{DBLP:journals/corr/abs-1905-07961}.
Sample prefix notation as well as its corresponding TPTP statements can be seen in Table~\ref{t:miz-tptp}.

Our {\elaborator} takes a TPTP-THF pseudo-formula as input and returns a boolean value indicating whether the pseudo-formula can be elaborated or not.
To use the {\elaborator} we translate the TPTP-FOF prefix format into the TPTP-THF format by a series of Prolog programs and related scripts.
If the pseudo-formula can be elaborated, the {\elaborator} will also output the corresponding elaborated TPTP-FOF formula.
In that case, the original pseudo-formula is back-translated by the MPTP toolchain into a Mizar statement, forming the feedback loop.
Figure~\ref{fig:format} illustrates the transformation of the file formats in the feedback loop.

\subsection{Experiments}\label{ss:elab-exp}

\begin{table}[!ht]
	\renewcommand{\arraystretch}{1.2}
	\caption{Model evaluation on the Mizar-TPTP dataset}\label{t:eval-elab}
	\begin{tabular}{m{0.27\columnwidth}>{\centering\arraybackslash}m{0.10\columnwidth}>{\centering\arraybackslash}m{0.10\columnwidth}>{\centering\arraybackslash}m{0.10\columnwidth}>{\centering\arraybackslash}m{0.10\columnwidth}>{\centering\arraybackslash}m{0.10\columnwidth}}
		\toprule
		{\small 2000 samples} & Iter. 1 & Iter. 2 & Iter. 3 & Iter. 4 & Iter. 5 \\
		\midrule
		BLEU & 42.3 & 85.5 & 88.7 & 88.0 & 87.4 \\
		Perplexity & 2.68 & 1.27 & 1.14 & 1.18 & 1.17 \\
		Edit distance 0 & 2.05\% & 2.40\% & 2.00\% & 5.15\% & 4.45\% \\
		Edit distance $\leq$1 & 9.70\% & 13.4\% & 20.9\% & 21.6\% & 19.2\% \\
		Edit distance $\leq$2 & 22.3\% & 25.0\% & 38.7\% & 36.9\% & 34.6\% \\
		Edit distance $\leq$3 & 32.6\% & 34.3\% & 49.3\% & 48.5\% & 45.4\% \\
		\bottomrule
	\end{tabular}
\end{table}

\begin{figure*}[htb]
  \includegraphics[scale=0.5]{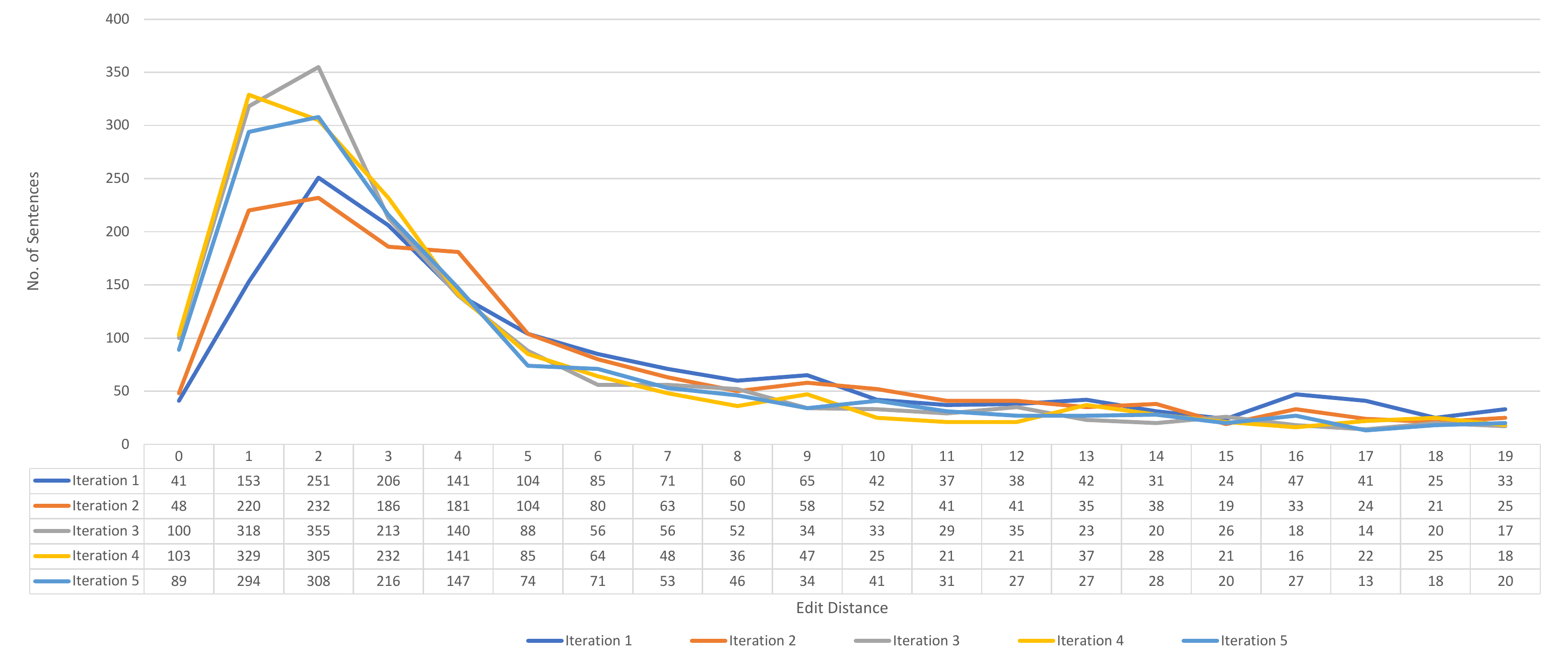}
  \caption{Distribution of edit distances in the elaboration experiment (total 2000 sentences)}
  \label{fig:distr-ed}
\end{figure*}

We evaluate the result of the data augmentation toolchain based on type elaboration by comparing the distribution of edit distances across the first five iterations of the feedback loop.

In Figure~\ref{fig:distr-ed} we can see that there is nearly an 100\% increase of values at edit distance 0 (exact match), indicating an improvement of translation quality.
The whole distribution curves in the first three iterations are becoming more skewed to the left, indicating a general improvement of translation quality toward exact matches.
However, as we can also see from Table~\ref{t:eval-elab}, starting from the fourth iteration the improvement is no longer significant.
The fourth iteration has the highest concentration of edit distance up to 2.
After that the distribution begins to drop and flatten.
This indicates that our trained model is beginning to overfit.
Such an overfit is likely to happen given limited training data.
This is unfortunate.
However, it is notable that from from Iteration 1 to Iteration 3, we have only had a 16.41\% increase in the size of the training data but we achieved a 100\% increase in the number of exact matches.
This shows that data augmentation based on strong semantic methods such as type elaboration can be very useful.

\section{Related Works}\label{s:related}
There are several works directly related to autoformalization of mathematics.
The earliest goes back to Wang (1954)~\cite{haowang} which hinted at the potential automation of formalization of mathematics.
In 1994, a group of anonymous authors declared the \textit{QED Manifesto}~\cite{qed-manifesto} which set out the vision of formalization of all mathematical knowledge in a computer-based database.
Zinn (2004)~\cite{Zinn04understandinginformal} used domain discourse theory to parse a number theory textbook.
The Naproche system~\cite{naproche} invented a controlled natural language that looks very similar to normal natural language text but compilable to proof obligations for automated theorem prover.
The vision set out in the QED Manifesto is exactly what the authors are trying to achieve.
It is our hope that the recent advent of neural network-based machine learning can finally provide a push to its realization.

\section{Conclusion}\label{s:concl}
In this paper we have proposed autoformalization of mathematics and performed first
experiments using unsupervised machine learning for translating LaTeX to formal proof
assistant statements.
We have also proposed a soft type elaboration
mechanism for data augmentation and integrated it into the supervised learning
framework for autoformalization. We practically evaluated various approaches and showed the results.
We think this is a promising direction and envision the possible future of the formalization community and the AI community embracing each other.

There are several directions that are worthy of further investigation: a primary concern is
collecting more high-quality data and conducting more comprehensive experiments.
It will be good if we can merge the ideas from new models and quickly test them in a single pipeline.
This could help us push to the extreme the capability of neural machine translation on our informal-to-formal datasets.

To adapt the existing models the format of our datasets is based on sequences only.
We have so far not experimented with the inclusion of other features such as the syntax tree of a formula or the required environment declaration of a statement.
This enriched information could have the potential to enrich the translation quality.
It could also open possibility for us to explore other non-sequential input formats.

As we have seen in Section \ref{s:exp}, even in unsupervised learning, it is still important to obtain a sizable aligned datasets for a reasonable evaluation.
To further harness the power of neural machine translation, we are currently still bound by the limitation of relevant datasets available.
Looking into the informal world, the amount of mathematics literature is actually quite abundant, but we still lack a way to combine those with the ITP knowhow we have at hand.
We believe that this is the major issue to be solved in autoformalization of mathematics.
It will be the long-term focus of our continual research effort.
At the current stage, we are exploring the possibility of joint embedding of multiple proof assistant libraries to see for opportunity to increase the size of our formal data.

\section*{Acknowledgement}
This work is supported by ERC grant no. 714034 \textit{SMART} and ERC grant no. 649043 \textit{AI4REASON}, by the Czech project
AI\&Reasoning CZ.02.1.01/0.0/0.0/15\_003/0000466, and the European Regional Development Fund.
We thank the anonymous reviewers for their helpful comments on improving the presentation of this paper.

\bibliographystyle{ACM-Reference-Format}
\bibliography{biblio}

\end{document}